\documentclass[twocolumn,pra,aps,showpacs,superscriptaddress,footinbib]{revtex4-1}

\usepackage[english]{babel}
\usepackage[utf8]{inputenc}

\usepackage{url,psfrag,graphicx}
\usepackage{dcolumn}
\usepackage{amsmath,amssymb,amsthm}
\usepackage{bm}
\usepackage{pstricks}
\usepackage{hyperref}
\usepackage{epsfig}
\usepackage{verbatim}
\usepackage{placeins}
\def\ket|#1>{| #1 \rangle}
\def\bra<#1|{\langle #1 |}
\def\<{\left\langle}
\def\>{\right\rangle}
\def\{{\left\lbrace}
\def\}{\right\rbrace}
\def\({\left(}
\def\){\right)}
\def\[{\left[}
  \def\]{\right]}
\def\nn{\nonumber}
\def\beq{\begin{equation}}
\def\eeq{\end{equation}}
\def\Var{\textrm{Var}}
\def\CM{\text{cm}}

\begin{document}

\title{Universal fluctuations of global geometrical measurements in planar
  clusters}

\author{Silvia N. Santalla}
\affiliation{Departamento de Física \&\ GISC, Universidad Carlos III
  de Madrid, Leganés (Spain)} 

\author{Iván Álvarez Domenech}
\affiliation{Departamento de Física Matemática y de Fluidos, UNED,
  Madrid (Spain)}

\author{Daniel Villarrubia}
\affiliation{Departamento de Matemáticas \&\ GISC, Universidad Carlos
  III de Madrid, Leganés (Spain)} 

\author{Rodolfo Cuerno}
\affiliation{Departamento de Matemáticas \&\ GISC, Universidad Carlos
  III de Madrid, Leganés (Spain)} 

\author{Javier Rodríguez-Laguna}
\affiliation{Departamento de Física Fundamental, UNED, Madrid (Spain)}

\begin{abstract}
We characterize universal features of the sample-to-sample
fluctuations of global geometrical observables, such as the area, width, length, 
and center-of-mass position, in random growing
planar clusters. Our examples are taken from simulations of both continuous
and discrete models of kinetically rough interfaces, including several
universality classes, such as
Kardar-Parisi-Zhang. We mostly focus on the scaling behavior 
with time of the sample-to-sample deviation for those global magnitudes, but we
have also characterized their histograms and correlations.
\end{abstract}

\date{November 26, 2023}

\maketitle


\section{Introduction}

Characterizing the statistical properties of rough interfaces away
from equilibrium is one of the main tasks in a variety of scientific contexts, 
such as the growth of solid phases in contact with a
vapour \cite{Barabasi}, liquid-crystal turbulence \cite{Takeuchi.11},
the shapes of isochrone curves on rough terrains
\cite{Santalla.15,Cordoba.18}, the growth of bacterial colonies or
cell aggregates \cite{Santalla.18}, or even the shape of a city skyline
\cite{Najem.20}. One of the most relevant insights was provided by the
Family-Vicsek (FV) dynamic scaling Ansatz \cite{Family.85}, which proposed that the
width, or roughness, of a rough interface grows with time $t$ as a power-law, 
$W\sim t^\beta$, where $\beta$ is called the {\em growth exponent}, up to a
saturation $t_{\text{sat}}\sim L^z$, where $z$ is called the {\em
dynamical exponent} and $L$ is the lateral size of the system. 
The FV Ansatz suggests the existence of a
well-defined {\em correlation length}, $\xi \sim t^{1/z}$, such that
the roughness at length-scales $\ell\ll\xi$ will always be saturated,
$w(\ell) \sim \ell^\alpha$, where $\alpha$ is the 
{\em roughness exponent}, the three exponents being related as $\alpha=\beta z$
within the FV formalism \cite{Barabasi,Halpin.15}.

The values of the scaling exponents $\beta$ and $z$ are typical
hallmarks of the kinetic roughening {\em universality class}. For example, 
for one-dimensional (1D) interfaces, $\beta=1/3$ and $z=3/2$ in the 
Kardar-Parisi-Zhang (KPZ) universality class \cite{Kardar.86,Barabasi,Halpin.15}, which 
is associated with growth along the local normal direction combined
with surface tension effects and time-dependent noise. Interestingly, the
KPZ universality class is able to fix also the one-point and the two-point 
(correlation) statistics of the local interface or front fluctuations, which 
are associated with Airy processes of different types, depending on whether 
the overall symmetry of the growth system is e.g.\ flat or circular 
\cite{Praehofer.02,Takeuchi.11,Corwin.13,Halpin.15}. Additionally, the statistical 
properties of global system quantities like the (squared) roughness $W^2$ has 
been characterized in detail for globally flat KPZ interfaces (the case 
for e.g.\ periodic boundary conditions) \cite{Foltin.94,Antal.02,Halpin.15}. 
Notably, an equivalent result seems to be lacking for the case of growing 
two-dimensional clusters, which in general remains somewhat less understood, 
in spite of its large interest for diverse contexts from epitaxial 
growth \cite{Misbah.10} to cellular aggregates \cite{Muzzio.16}. For instance, 
as clarified in Ref.\ \cite{Ferreira.06}, the additional degrees of freedom 
implied by the dynamics of 2D clusters (like the evolution of their center of 
mass) has sometimes even led to  incorrect identification of exponent values 
and universality classes for their  corresponding 1D fronts. More recently \cite{Saito.12}, 
suitable characterizations of the cluster dynamics has been shown to extract 
correct exponent values and even the detailed time evolution for certain 
measures of 2D clusters under growth or dissolution conditions.

The aim of the present article is to characterize global properties defined
in each case as a whole for 2D growing clusters. Through a scaling analysis, 
non-trivial predictions, like scaling exponent values, will be derived from general 
considerations on the sample-to-sample fluctuations of such properties. Specifically, 
we will  consider the average radius $R$, the total area $A$, the total width $W$, the 
(suitably regularized) length $L$, and the center-of-mass displacement,
$R_{\text{cm}}$. As we will show, the expected values of these magnitudes and 
their deviations grow as power laws of time, with exponents which depend on the values
of $\beta$ and $z$. Previous attempts to predict the sample-to-sample
fluctuations of global variables have been made in the past. For
example, the center-of-mass displacement was predicted to grow as
$t^{1/6}$ in KPZ clusters \cite{Saito.12}, as we here confirm for some additional
examples. Moreover, we will also describe the correlation between these global
magnitudes and their full histograms, which in some cases is Gaussian,
and for $R_{\text{cm}}^2$ we will show that it corresponds to a
$\chi^2$-distribution. The case of the squared global roughness $W^2$, which has 
been extensively studied for globally flat interfaces 
\cite{Foltin.94,Antal.02,Halpin.15}, is more involved, but seems to share some 
similarities with its flat counterpart.

\begin{figure}
  \hbox{\includegraphics[width=4cm]{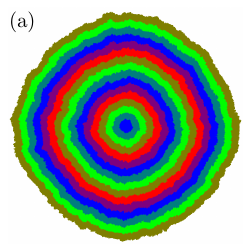}
  \hskip 0.5cm
    \includegraphics[width=4cm]{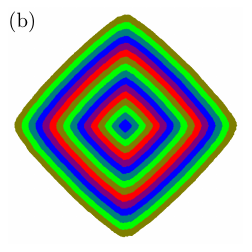}}
  \hbox{\includegraphics[width=4cm]{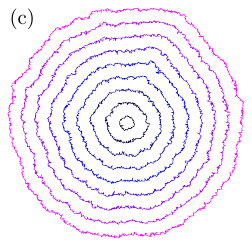}
  \hskip 0.5cm
    \includegraphics[width=4cm]{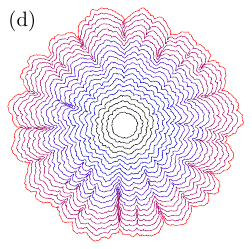}}
  \caption{Typical profiles for the different physical examples
    considered in this article (see definitions in 
    Sec.\ \ref{sec:results}). In all cases, different colors
    represent different growth times. (a) FPP ball with $d_c=1$;
    (b) FPP ball with $d_c=20$; (c) isochrone curves in the random 
    metrics problem; (d) consecutive snapshots of the time evolution of
    the interface in the shadowing model of bacterial colony growth.}
  \label{fig:profiles}
\end{figure}

We will apply our scaling estimates to simulations of growing planar clusters generated
by different physical systems, whose interfaces (boundaries) are known to follow 
FV scaling. We will start by discussing neighborhoods (balls) in the {\em first-passage percolation}
(FPP) model, whose boundaries present 1D KPZ universality in the asymptotic regime
\cite{Hammersley.65,Cordoba.18} if discrete lattice effects are suitably
taken into account \cite{Alvarez.23}. Typical FPP balls are shown in
Fig.\ \ref{fig:profiles} (a) and (b), depending on the level of
disorder. The continuous analogue of FPP is called the {\em random
metric problem}, where we consider isochrone curves on a
two-dimensional manifold with a random (disordered) metric field which is flat on
average, with only short-range correlations \cite{Santalla.15}.
Typical isochrones are shown in Fig.\ \ref{fig:profiles} (c).
Interestingly, different types of one-point and two-point
correlation functions are obtained depending on whether the underlying manifold is
a plane, a cone or a cylinder \cite{Santalla.17}, although all these cases belong to the 1D
KPZ universality class.

Additional interesting examples of rough interfaces with overall circular geometry 
are provided by the fronts of growing bacterial colonies \cite{BenJacob.00}, for which the most relevant
physical parameters are the motility and the nutrient concentration.
For many values of these parameters 1D KPZ scaling can be observed, but 
other behaviors are also possible. Specifically, for low motility and low nutrient
concentration, {\em shadowing effects} ---whereby the growth rate at each interface 
point depends on the angle under which the exterior of the cluster can be seen \cite{Santalla.18}--- 
dominate the interfacial dynamics. In Fig.\ \ref{fig:profiles}
(d) we show a typical time evolution for an interface described by this shadowing model.

This article is organized as follows. In Sec.\ \ref{sec:theory} we
discuss our theoretical framework in order to determine the scaling
exponents for global magnitudes of clusters following FV scaling.
Our predictions are then tested on FPP clusters, random metrics
isochrones, and bacterial colonies in Sec.\ \ref{sec:results}. Additional 
results for the histograms and the correlations
between magnitudes, are reported in Sec.\ \ref{sec:other}. The article
ends with a summary of our conclusions and some proposals for further
work.

\section{Fluctuations of global geometrical observables}
\label{sec:theory}

Let us consider a growing planar cluster whose boundary is described 
by a polar curve advancing in time, $r(\theta,t)$, subject to a stochastic 
evolution law which we may assume (along with the initial condition) to be 
isotropic. Let us also consider a local observable $u$ which is a
function of $r$, of $r'\equiv dr/d\theta$, and of $\theta$ itself, which we
will denote by $u(r(\theta),r'(\theta),\theta)$ or just $u(\theta)$
for short when it is convenient. Its two-point correlation function
can be defined as

\beq
C_u(\hat\theta)\equiv
\<u(\theta)u(\theta+\hat\theta)\>-\<u(\theta)\>\<u(\theta+\hat\theta)\>,
\label{eq:def_corr}
\eeq
where we will denote the angular distance between the two points by
$\hat\theta$. In the asymptotic regime, we assume the following
scaling form for the correlation function,

\beq
C_u(\hat\theta) \approx A_u t^{2\phi_u} g_2(B_u t^{1-\zeta} \hat\theta),
\label{eq:corr_form}
\eeq
where $\phi_u$ is the corresponding scaling exponent, $A_u$ and $B_u$ are
constants, $\zeta=1/z$ is the inverse of the dynamical exponent, and
$g_2(x)$ is a continuous function such that $g(x)\sim 1$ for
$x\ll 1$, and $g(x)\to 0$ sufficiently fast for $x\to\infty$.

The reason behind the form of Eq.\ \eqref{eq:corr_form} is as follows.
Assuming that the expected value of the radius grows as $\<r\>\sim t$
and that the correlation length grows as $\xi \sim t^\zeta$, as it is
the case in the Family-Vicsek Ansatz \cite{Family.85}, then the
angular aperture of each correlated patch along the front will be $\delta\theta_0 \sim
t^{\zeta-1}$. Then, the argument of the $g_2(x)$ function should be
$\delta\theta/\delta\theta_0$, as shown in Eq. \eqref{eq:corr_form}.

\medskip

As a first example, let us consider a cluster family corresponding to
the KPZ universality class \cite{Kardar.86} and the local observable
$u=r$. In that case, we have $\phi_u=\beta=1/3$, and
$C_r(\hat\theta)\sim t^{2\beta} g_2(B_rt^{1-\zeta} \hat\theta)$, where
$\zeta=1/z=2/3$. Assuming that our cluster ensemble possesses a
well-defined correlation length, it seems appropriate to assume that
all scaling observables will present a similar structure in their
correlation functions.

\medskip

Let us now consider the statistical distribution of the values of a
{\em global} measure of geometric origin, such as the {\em area} or
the {\em length}, which can be written as

\begin{equation}
U=\int_0^{2\pi} d\theta\; u(r,r',\theta),
\label{eq:u}
\end{equation}
This work is devoted to evaluating the sample-to-sample fluctuations
of any global measure $U$, which will be quantified through their
deviation, $\Delta U$, or their variance, 

\beq
\Var(U) \equiv (\Delta U)^2 \equiv \<U^2\>-\<U\>^2.
\eeq
The first two moments can be written as

\begin{align}
  \<U\>&=\int_0^{2\pi} d\theta\; \< u(\theta) \>, \nn\\
  \<U^2\>&=
  \int_0^{2\pi} d\theta \int_0^{2\pi} d\theta'\; \<u(\theta) u(\theta')\>,
\end{align}
thus allowing us to use a more compact notation for the variance of $U$,

\begin{align}
\Var(U) &=\int d\theta d\theta'\;
\<u(\theta)u(\theta')\>-\<u(\theta)\>\<u(\theta')\> \nonumber \\
&= \int d\theta d\theta'\; C_u(\theta-\theta'),
\label{eq:varcorr}
\end{align}
where $C_u(\theta-\theta')$ is again the correlation function for the
observable $u$, defined in Eq.\ \eqref{eq:def_corr}. Thus, we can
compute the variance of $U$:

\begin{align}
  \Var(U)&=\int d\theta d\theta'\; C_u(\hat\theta)
  = 2\pi \int_0^{2\pi} d\hat\theta\; C_u(\hat\theta) \nonumber\\
  &= 2\pi A_u t^{2\phi_u} \int_0^{2\pi} d\hat\theta\;
  g_2(B_u t^{1-\zeta} \hat\theta) \nonumber\\
  &\approx {2\pi A_u\over B_u} t^{2\phi_u+\zeta-1} \int_0^\infty dx\; g_2(x).
\end{align}
Assuming that $g_2(x)$ decays fast enough for large values of its
argument, the last integral is finite and does not affect the scaling behavior, 
thus leading to an estimate for the sample-to-sample deviation of $U$,

\beq
\Delta U \sim t^{\phi_u + (\zeta-1)/2}.
\label{eq:scaling}
\eeq
This expression can be motivated in a heuristic way as follows.
The variance of the average of $N$ independent identically distributed
(i.i.d.) random variables $\{u_i\}$ is $\Var(\bar u)=\Var(u)/N$. Yet,
if these random variables are strongly correlated among themselves,
with $N_P$ independent groups, then it is straightforward to prove
that $\Var(\bar u)=\Var(u)/N_P$. If the system radius grows approximately 
as $t$ and the correlation length grows as $t^\zeta$, then each profile
possesses $n_P\sim t^{1-\zeta}$ independent patches. Therefore, the
variance of a global variable $U$ must be given by

\beq
\Delta U \sim {\Delta u \over \sqrt{n_P}} \sim t^{\phi_u+(\zeta-1)/2},
\eeq
which coincides with the result shown in Eq. \eqref{eq:scaling}.

\bigskip

The rest of this section is devoted to the theoretical analysis of the
sample-to-sample fluctuations of several global geometrical
observables, such as the (average) radius, area, width, center of mass
position, and interfacial length.

\subsection{Radius}

As it has been discussed above, the sample-to-sample fluctuations of
the average radius of the cluster,

\beq
R = \int_0^{2\pi} d\theta\; {r \over 2\pi},
\label{eq:radius}
\eeq
can be obtained by applying our formalism to the observable
$u(r,r',\theta)=r$, which has the associated scaling exponent
$\phi_r=\beta$, thus yielding the prediction $\Delta R \sim
t^{\beta+(\zeta-1)/2}$. For example, in the 1D KPZ case, $\Delta R
\sim t^{1/6}$, which has been numerically verified for balls in random
metrics \cite{Santalla.15}.

\subsection{Area}

Let us now consider the cluster area, which is given by

\beq
A=\int_0^{2\pi} d\theta\; {r^2\over 2}.
\eeq
Within our formalism, its sample-to-sample fluctuations can be
obtained choosing $u(r,r',\theta)=r^2$. The associated scaling
exponent can be found through classical uncertainty propagation,
$\Delta(r^2) \sim r\Delta r \sim t^{1+\beta}$. Thus, $\Delta A\sim
t^{\beta+(\zeta+1)/2}$. For 1D KPZ, our prediction is $\Delta A \sim
t^{7/6}$.

\subsection{Width}

In our next example we will consider the sample-to-sample fluctuations
of the interface width, defined as

\beq
W^2=\int_0^{2\pi} d\theta\; \frac{(r-R)^2}{2\pi},
\eeq
so that the fluctuations in $W^2$ can be obtained using our rule. The
integrand $(r-R)^2$ has fluctuations of order $t^{2\beta}$. Thus, its
variance scales as $t^{4\beta}$, and we have

\beq
\Var(W^2) \sim t^{4\beta-1+\zeta}, 
\eeq
which yields $\Delta(W^2) \sim t^{2\beta-1/2+\zeta/2}$. Yet, we have
$\Delta(W^2) \sim W\Delta W$ and $W\sim t^\beta$, leading us to
predict

\beq
\Delta W\sim t^{\beta-1/2+\zeta/2}.
\eeq
For example, in 1D KPZ, we have $\Delta W \sim t^{1/6}$.

\subsection{Center-of-mass position}

In absence of fluctuations, the center-of-mass (CM) of a growing
cluster starting out as a tiny circle must remain at the origin. But,
even though the statistical properties of the cluster are isotropic,
each sample presents unbalances which will give rise to fluctuations
in the CM position \cite{Ferreira.06,Saito.12}, 

\begin{align}
X_\CM &= {1\over A} \int_0^{2\pi} d\theta\; {r^3\over
  2}\cos(\theta),\nn\\
Y_\CM &= {1\over A} \int_0^{2\pi} d\theta\; {r^3\over
  2}\sin(\theta).
\end{align}
Each of them presents an expectation value of zero, and a non-zero
variance, which shows up in the expected value of the squared
displacement,

\beq
R_\CM^2=X_\CM^2+Y_\CM^2 \geq 0.
\eeq
Let us evaluate the sample-to-sample fluctuations of the following
associated magnitude, which neglects the explicit angular dependence,

\beq
U = \int_0^{2\pi} d\theta\; {r^3\over 2}, 
\label{eq:rcm}
\eeq
and analyzing the local fluctuations of $u(r,r',\theta)=r^3/2$, i.e.
$\Delta u\sim t^{2+\beta}$. Thus, we have $\Var(U)\sim
t^{4+2\beta+\zeta-1}$. Now, we may guess that the scaling behavior of
$U$ is the same as that for $X_\CM A$ or $Y_\CM A$. Thus, employing
the usual uncertainty propagation techniques,

\beq
\Delta X_\CM \approx {\Delta U\over A}  + U {\Delta A\over A^2}.
\eeq
Both terms scale in the same way, as $t^{\beta+(1-\zeta)/2}$. Thus,

\beq
R_\CM^2 \sim t^{2\beta+\zeta-1}.
\eeq
Thus, the center of mass fluctuates with the same exponent as the
average radius. For 1D KPZ, this leads to $R_\CM\sim t^{1/6}$ \cite{Saito.12}.

\subsection{Length}

The length of a cluster, $L$, is a different type of observable. First
of all, its measure may depend on the UV-cutoff if the interface
presents a non-trivial fractal nature. Yet, we will assume that the
interface is always smooth at the microscopic level and that the total
length increases linearly in time, $L\sim t$. If the (radial) slopes are small,
i.e. $r'/r\ll 1$, we may write,

\begin{align}
L &= \int_0^{2\pi} d\theta\; \sqrt{r^2+r'^2} = \int_0^{2\pi} d\theta\; r
\sqrt{1+\({r'\over r}\)^2} \nonumber \\ 
&\approx \int_0^{2\pi} d\theta\; \(r  + {1\over 2} {r'^2\over r}  \),
\end{align}
which forces us to consider the fluctuations of the local derivative of
the radius with respect to the angle, $r'(\theta)$. In order to do
that, let us consider a small angle difference, $\delta\theta$, and
evaluate

\beq
r'\approx {r(\theta+\delta\theta)-r(\theta)\over \delta\theta},
\eeq
so we have

\begin{align}
\Var(r') =&\<r'^2\>-\<r'\>^2={2\over\delta\theta^2} \[
\Var(r)-C_r(\delta\theta) \] \nn\\
=&
{2t^{2\beta}\over\delta\theta^2} \[g_2(0)-g_2(t^{1-\zeta}\delta\theta)\]
\nonumber \\
\approx &-2g_2'(0) {t^{2\beta+1-\zeta}\over \delta\theta}
- g_2''(0)t^{2(\beta+1-\zeta)}.
\label{eq:scaling_rp}
\end{align}
The first term in Eq. \eqref{eq:scaling_rp} diverges as $\delta\theta
\to 0^+$ unless we can ensure $g_2'(0)=0$, which seems a reasonable
assumption within our framework. In that case, the second
term provides the complete scaling,

\beq
\Delta(r') \sim t^{\beta-\zeta+1},
\label{eq:scaling_rprime}
\eeq
which becomes $\Delta(r')\sim t^{2/3}$ in the 1D KPZ case. Indeed,
we have $r'/r\sim t^{-1/3}$, so this ratio becomes negliglibly small for large
times, as expected.

Let us provide an intuitive explanation for this scaling form. Once we
have ensured that the interface is smooth, we may estimate the
derivative $r'$ by assuming that the radii will span the full range of
$W$ within each correlated patch of size $\xi$. Thus, we expect
$\Delta(r')\sim \Delta r/\delta \theta_0 \sim W/(\xi/R) \sim
t^{\beta-\zeta+1}$.

The scaling form for the slopes allows us to evaluate the
sample-to-sample fluctuations of the cluster length, employing
Eq.\ \eqref{eq:scaling}. Indeed,

\beq
\Var(L) \sim t^{2(\beta-\zeta+1)} t^{\zeta-1} = t^{2\beta-\zeta+1},
\label{eq:length}
\eeq
which for 1D KPZ is just $\Delta L \sim t^{1/2}$.

\medskip

As a curiosity, we may define the length-to-radius ratio of any
cluster family, or the generalized value of $2\hat\pi$. Of course,
this $2\hat\pi$ value may in general depend on the measurement scale
if the interface is fractal, but assuming a smooth behavior below the
UV-cutoff, we may describe the sample-to-sample fluctuations of the
$\hat\pi$ value for the 1D KPZ case,

\begin{equation}
  \Delta(2\hat\pi)
  \sim {\Delta L\over R} + L {\Delta R\over R^2},
\label{eq:fluct_2pi_1}
\end{equation}
The first term scales as $t^{\beta-\zeta/2-1/2}$, while the second
one scales as $t^{\beta+\zeta/2-3/2}$. The first term will be dominant
whenever $\zeta\leq 1$, which is the case in all the considered
universality classes. Therefore, we may conjecture that

\beq
\Delta(2\hat\pi) \sim t^{\beta-\zeta/2-1/2},
\label{eq:fluct_2pi}
\eeq
which for 1D KPZ leads to $\Delta(2\hat\pi) \sim t^{-1/2}$.
Therefore, we see that the length-to-radius ratio of different samples
will converge to a common value in the long run.

\subsection{Summary of scaling predictions}

The theoretical predictions from the scaling analysis discussed in this
section are summarized in Table \ref{tab:scaling}, which shows the scaling 
exponent with time for the sample-to-sample fluctuations of each observable; 
for the sake of reference, the exponent values for the 1D KPZ are collected 
in the last column.

\begin{table}[h]
 \begin{tabular}{c|c|c|c|}
      Observable & Average &  Fluctuations & 1D KPZ\\
      \hline
      $R$ & $1$ & {$\beta+(\zeta-1)/2$} & 1/6\\
       $A$ & $2$ & {$\beta+(\zeta+1)/2$}& 7/6\\
      $W$ & $\beta$ & {$\beta+(\zeta-1)/2$}& 1/6 \\
      $R_\CM$    & {$\beta+(\zeta-1)/2$} & $-$& 1/6\\
      $L$ & $1$ & {$\beta-(\zeta-1)/2$}& 1/2\\
      \hline
    \end{tabular}
    \caption{Scaling behavior for the sample-to-sample average and deviation of different global geometrical observables, and expected numerical values of the new exponents for the 1D KPZ universality class, $\beta=1/3$ and $\zeta=2/3$.}
    \label{tab:scaling}
 \end{table}

Thus, the following predictions can be made:

\begin{itemize}
  \item The scaling exponent values for the sample-to-sample variation of
    the average radius, the width, and the CM displacement coincide.
  \item The scaling exponent for the sample-to-sample variation of the
    area equals the previous exponent plus one.
  \item We may obtain both the growth and the dynamical exponents
    using (e.g.) the fluctuations of the average radius and the
    interface length.
\end{itemize}


\begin{figure}
  \includegraphics[width=8cm]{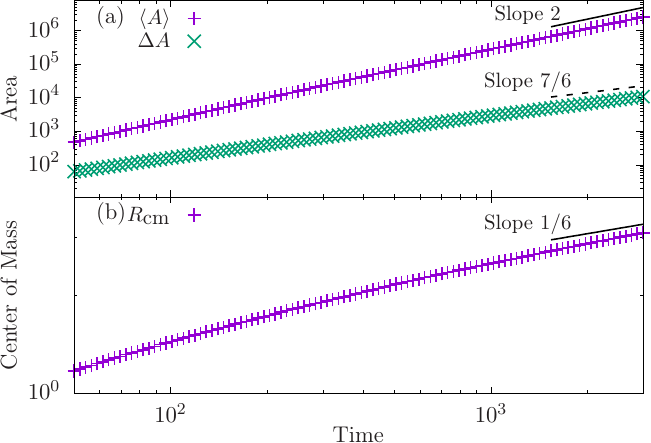}
  \caption{Deviation of some global magnitudes for FPP isochrones with
    $d_c=1$: average radius, area, and center-of-mass deviation. The straight lines in each panel represent the corresponding theoretical 
    expectation for 1D KPZ behavior, see Table \ref{tab:scaling}.}
  \label{fig:fpp_1}
\end{figure}

\begin{figure}
  \includegraphics[width=8cm]{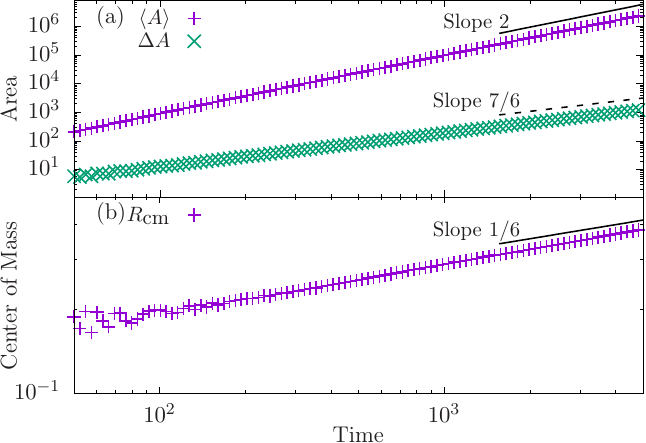}
  \caption{Deviation of some global magnitudes for FPP isochrones with
    $d_c=20$: average radius, area, and center-of-mass deviation. 
    The straight lines in each panel represent the corresponding theoretical 
    expectation for 1D KPZ behavior, see Table \ref{tab:scaling}.}
  \label{fig:fpp_2}
\end{figure}

\section{Numerical results}
\label{sec:results}

In this section we will compare our theoretical predictions, collected
in Table \ref{tab:scaling}, with numerical simulations of different
models which are known to follow FV scaling. We will discuss
first-passage percolation (FPP), random metrics, and shadowing models.
Besides these models, we have also performed simulations on two
flavors of the random deposition model \cite{Barabasi} for circular
clusters, which are shown in Appendix \ref{sec:rd}.

\subsection{First-passage percolation} 
\label{sec:fpp}

Our first example will be first-passage percolation (FPP) on a square
$L\times L$ lattice, which is defined as follows. Each lattice
link $k$ has an associated crossing-time, $\{t_k\}$, which are
independently identically distributed (i.i.d.) random variables
extracted from a certain probability distribution, with cumulative
probability function $F(t)$ such that $F(0)=0$. Employing Dijkstra's
algorithm \cite{Cormen.90} we find the minimal arrival time at every
vertex $i$ starting from the lattice center \cite{Cordoba.18},
$\{T_i\}$. Then we determine the {\em ball} of radius $R$ as the set
of vertices for which $T_i\leq R$. We have chosen uniform
distributions for the crossing-times with mean $\mu$ and deviation
$\sigma$. The balls are then characterized by the {\em crossover
  length} $d_c\equiv \mu^2/(3\sigma^2)$. In Fig.\ \ref{fig:profiles}
(a) and (b) we can see typical profiles using $d_c\approx 1$ and
$d_c=20$. Notice that the average shape is nearly circular in the
first case, and similar to a diamond in the second. Yet, the
fluctuations are known to correspond to KPZ for all values of $d_c$
\cite{Villarrubia.20,Alvarez.23}.

We have run $10^4$ simulations on $L=2401$ FPP lattices, using uniform
time distributions with $\mu=5$ and different $d_c$. The time
evolution of the average and deviation of the area and the CM
displacement are shown in Fig.\ \ref{fig:fpp_1} for $d_c=1.04$ and in Fig.\
\ref{fig:fpp_2} for $d_c=20$. Notice that for this discrete model in
particular, the aforementioned global magnitudes are easier to obtain,
because they are measured in the bulk. The boundaries of the balls,
which are called {\em isochrone curves} or isochrones, present some subtle points
\cite{Alvarez.23} and have been left out from our present numerical study.
The solid black lines show the theoretical predictions, extracted
from the last column of Table \ref{tab:scaling} (1D KPZ behavior), and show good 
agreement with the simulation data.

\begin{figure}
  \includegraphics[width=8cm]{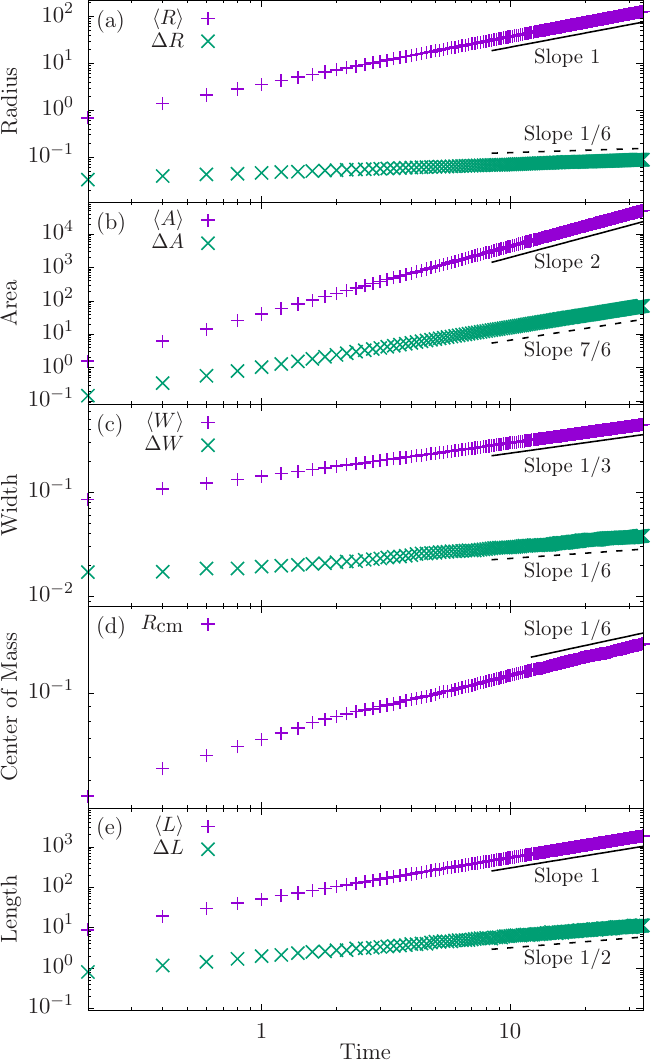}
  \caption{Expected values and deviations for the global magnitudes
    for random-metric isochrones: (a) Radius, (b) area, (c) width, (d)
    center-of-mass displacement, and (e) length.
    The straight lines in each panel represent the corresponding theoretical 
    expectation for 1D KPZ behavior, see Table \ref{tab:scaling}.}
  \label{fig:rm}
\end{figure}

\subsection{Random metrics}

The FPP problem is a discrete analogue of the more general {\em random
metrics} problem \cite{Santalla.15}. In the latter, we consider a random
two-dimensional manifold, flat in average, whose metric
tensor presents only short-distance correlations, and obtain the
isochrone curves by integrating Huygens' equation,

\beq
\partial_t \vec r = \vec n_g(\vec r),
\label{eq:Huygens}
\eeq
where $\vec n_g(\vec r)$ denotes the local normal to the isochrone at
position $\vec r$, according to the metric tensor $g$. Both the
isochrones and the times-of-arrival present very accurate 1D KPZ scaling
from the beginning \cite{Santalla.15}.

We have performed 1280 simulations of Huygens' equation, Eq.\
\eqref{eq:Huygens}. Each simulation starts out with a very small ball, with
initial radius $0.05$, and propagates it through a random metric field with
uniformly distributed eigenvalues $\lambda \in [1/20, 1]$, using a
time-step $\Delta t= 5 \times 10^{-3}$. We have obtained the full set
of global observables: average radius, area, width, CM displacement, and
length, whose time evolutions are shown in Fig.\ \ref{fig:rm}, along
with the theoretical predictions extracted from Table
\ref{tab:scaling}. Notice that, in all the considered cases, the
theoretical lines accurately describe the simulation data.

\begin{figure}
  \includegraphics[width=8cm]{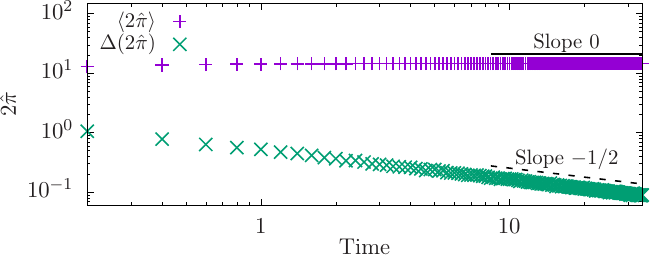}
  \caption{Expected values and variances of the length-to-radius
    ratio, or $2\hat\pi$, for random-metric isochrones, along with the 
    comparison to the theoretical prediction, Eq.\ \eqref{eq:fluct_2pi}, 
    for 1D KPZ behavior.}
  \label{fig:2pi}
\end{figure}

Furthermore, we have checked the length-to-radius ratio, i.e.\ the
value of $2\hat\pi$, and the results are shown in Fig.\ \ref{fig:2pi}.
Indeed, the theoretical predictions are once more correct, with the
ratio converging to a fixed value whose fluctuations decay as
$t^{-1/2}$, as predicted by Eq.\ \eqref{eq:fluct_2pi}.

\subsection{Shadowing model}

\begin{figure}
  \includegraphics[width=8cm]{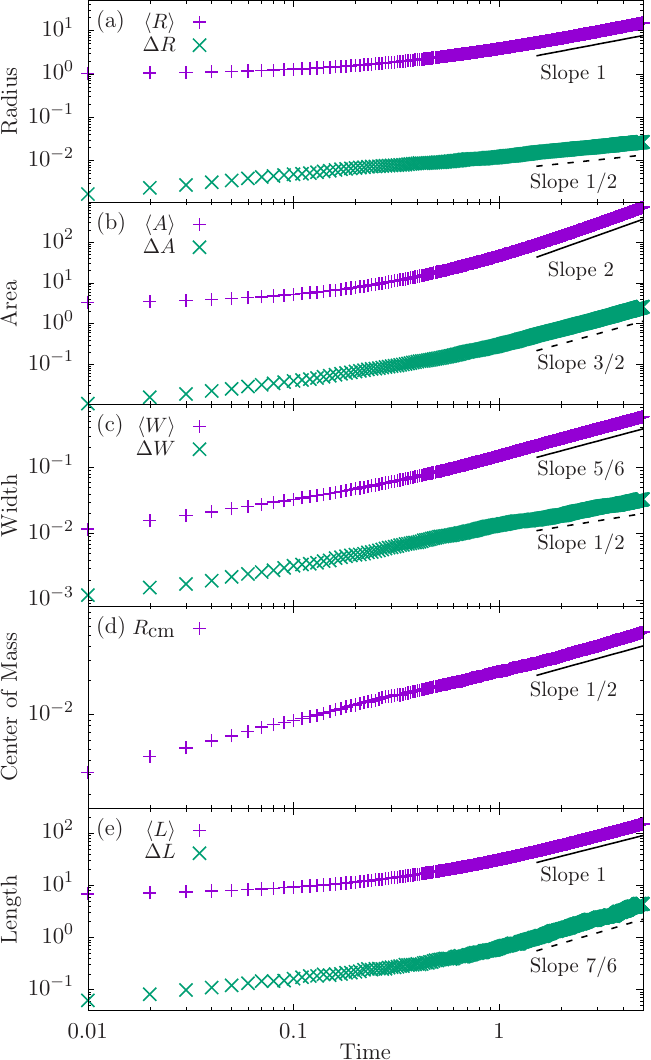}
    \caption{Expected values and deviations for the global magnitudes
      for the shadowing model: (a) Radius; (b) Area; (c) Width; (d)
      Center-of-mass displacement; (e) Length.}
  \label{fig:bact}
\end{figure}
Motivated by the morphological instabilities of the fronts of growing
planar bacterial colonies, shadowing effects may be introduced
phenomenologically into the radial KPZ equation, such that each point
at the interface moves along the normal direction with a velocity 
which is proportional to the local {\em aperture angle}, i.e.\ the fraction 
of rays emanating from the point which do
not intersect the interface \cite{Santalla.18,Santalla.21}. The
resulting continuum model is given by the following equation,

\begin{equation}
  \partial_t \vec r =
  \( A_0 + A_1 K(\vec r) + A_a \Theta_a(\vec r) + A_n \eta \) \vec n ,
\label{eq:bacterial_growth}
\end{equation}
where $\vec r$ is any interface point, $\vec n$ is the local exterior
normal, $K(\vec r)$ denotes the curvature of the interface at that
point, $\Theta_a(\vec r)$ is the local aperture angle, and $\eta$ is a
zero-average and unit variance, Gaussian, uncorrelated space-time
noise. Furthermore, $A_0$, $A_1$, $A_a$, and $A_n$ are positive
parameters which quantify, respectively, the relative strengths of the
average growth velocity of a planar front, surface tension, the
dependence on the aperture angle, and fluctuations.
For a convex smooth shape, the aperture angle is uniformly equal to
$\pi$. The subsequent dynamics tends to make peaks grow faster than
valleys, thus giving rise to morphological instabilities. Indeed, in
the long run the typical cluster is composed of a set of correlated
{\em lobes} separated by deep crevices whose angular distance is
nearly constant in time. An example can be seen in Fig.
\ref{fig:profiles} (d).

We have performed 500 simulations of Eq.\ \eqref{eq:bacterial_growth} using the 
same numerical scheme as in Ref.\ \cite{Santalla.18},
for initial radius 1, $A_0=0$, $A_1=0.1$, $A_a=1$, $A_n=0.1$, and time-step $\Delta t=10^{-4}$.
We have measured the full set of global magnitudes: average radius, area,
width, CM displacement, and length. The time evolution of their
average and deviation can be found in Fig.\ \ref{fig:bact}. Previous
work \cite{Santalla.18,Santalla.21} was able to unambiguously rule out
KPZ scaling for this model, even finding traces of nonuniversality
both in experiments and in simulations, and very precise values of the
critical exponents could not be ascertained for the FV behavior that
could nevertheless be confirmed. Yet, our present global measurements
agree with the scaling behavior predicted in Sec.\ \ref{sec:theory},
compatible with (non-KPZ) values for the scaling exponents
$\beta\approx 5/6$ and $\zeta=1/z\approx 1/3$, implying $\alpha=\beta
z\approx 5/2$.


\section{Other statistical properties}
\label{sec:other}

\begin{figure}
  \includegraphics[width=8cm]{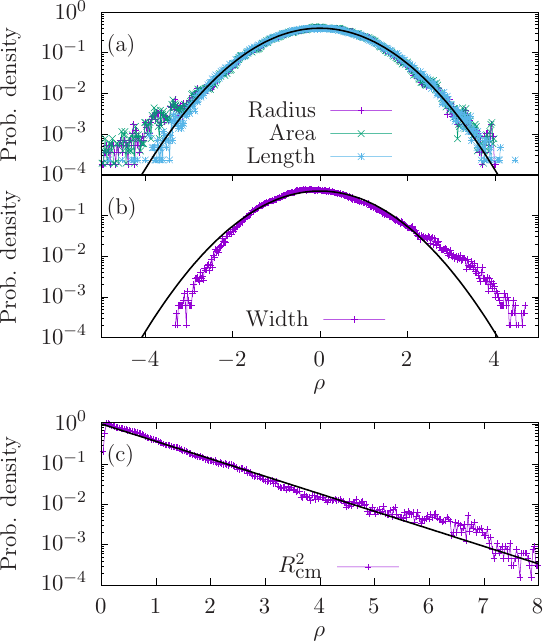}
  \caption{Probability density for the standardized global magnitudes, Eq.\,\eqref{eq:standGM}, considered
    for the random metrics system; (a) Radius, area and length, along
    with the unit Gaussian; (b) Squared width, along with an
    exponential decay, as predicted in other cases in the literature
    \cite{Foltin.94,Halpin.15}; (c) Squared CM displacement,
    $R_\CM^2$, along with the $\chi^2$-distribution for $k=2$ degrees
    of freedom (i.e. an exponential decay).}
  \label{fig:histog}
\end{figure}

\subsection{Histograms}

We may also provide some predictions for the full histograms of some
of the global observables considered. Indeed, if the number of patches
$n_P$ is large, the average radius, the area, and the length can be
considered to be the sum or average of a series of i.i.d.\ random
variables. The central limit theorem then predicts that, under very
broad circumstances, the probability distribution for the global
observables must be Gaussian. In the random metrics case, we have
considered the full set of values of the average radius, area, and
length, for a given time $t$, substracted their (time-dependent)
average and divided by their (time-dependent) deviation so that their
average becomes zero and their variance becomes one, i.e., we have
defined

\beq
\rho_i = {U_i - \<U(t)\> \over \sigma_U(t)}.
\label{eq:standGM}
\eeq
Then we have plotted the histograms of the full set of values
$\rho_i$ in Fig.\ \ref{fig:histog} (a), along with the unit Gaussian,
showing their correspondence. The prediction is specially good for the
length, with some deviation for the radius and area.

In Fig.\ \ref{fig:histog} (b) we show the histogram for the $\rho_i$
values corresponding to the square of the global width, which need not
be Gaussian. In fact, results associated to the KPZ class in band
geometry after saturation show a very skewed histogram with a
large-devations exponential decay \cite{Halpin.15}, which can be
accounted for by considering the behavior of random walks
\cite{Foltin.94,Antal.02}. Our case, which corresponds to circular
geometry and is not saturated, also shows a large-deviation
exponential decay, as we can see in Fig.\ \ref{fig:histog} (b).

Furthermore, Figure \ref{fig:histog} (c) shows the histogram for the
squared CM displacements, $R_\CM^2$, merely normalized to have
variance one. In this case, the theoretical prediction is not
Gaussian. Indeed, $R_\CM^2=X_\CM^2+Y_\CM^2$, where $X_\CM$ and $Y_\CM$
can be in turn considered to be Gaussian. Therefore, the sum of
squares must follow a $\chi^2$-distribution for two degrees of
freedom, which is an exponential distribution as we can indeed observe
in the plot.

\subsection{Correlations between global magnitudes}

\begin{figure}
  \includegraphics[width=8cm]{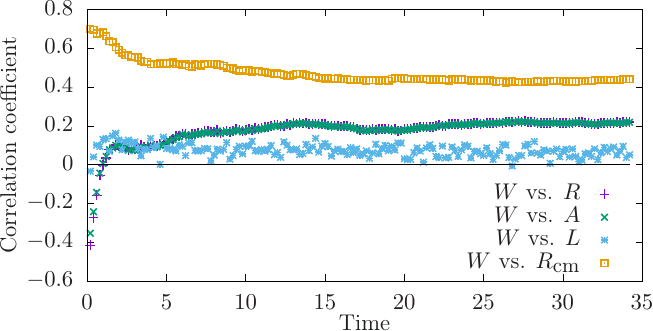}
  \caption{Correlation coefficients, Eq.\,\eqref{eq:ccoeff} between some pairs global
    magnitudes as a function of time, for the random metrics system:
    width vs radius, area, length, and center-of-mass displacement.}
  \label{fig:corr}
\end{figure}

It is interesting to consider whether the sample-to-sample
fluctuations of different global magnitudes present correlations.
Indeed, it is natural to expect that the fluctuations of the average
radius and the area must be strongly correlated, with a smaller (yet
positive) correlation between the CM displacement and the interface
width.

In order to obtain a theoretical prediction, we consider any two
global magnitudes $U$ and $V$, and define their correlation
coefficient,

\beq
r_{UV}\equiv {\<UV\>-\<U\>\<V\> \over \sigma_U \sigma_V}.
\label{eq:ccoeff}
\eeq
If we assume that the correlator between different magnitudes behaves
as [recall Eq.\ \eqref{eq:corr_form}]

\begin{align}
C_{uv}(\theta,\theta')
= A_{uv} t^{\phi_a+\phi_b} \; g_2(B_{uv} t^{1-\zeta} \hat\theta),
\end{align}
then

\begin{align}
\<UV\>-\<U\>\<V\> =& \int d\theta d\theta'\; C_{uv}(\theta,\theta')
\nn\\
\sim &
{2\pi A_{uv}\over B_{uv}} t^{\phi_u+\phi_v+\zeta-1} \int_0^\infty dx\; g_2(x),
\end{align}
whose time dependence is the same as that for the product
$\sigma_U\sigma_V$, thus concluding that the {\em correlation
coefficients approach time-independent values}. This is indeed what we
observe in Fig.\ \ref{fig:corr}, where we have considered the
correlation coefficients between the width and the other four global
observables in the random metrics simulations. Notice that the curves
for the average radius and the area overlap almost perfectly, because
the correlation coefficient between them is close to one.


\section{Conclusions}
\label{sec:conclusions}

In this work we have presented a scaling approach to the
sample-to-sample fluctuations of global geometrical observables
measured on random planar clusters whose fronts display statistical
properties satisfying the Family-Vicsek Ansatz. The chosen observables
were the average radius, the area, the width, the center-of-mass
displacement, and the length of the clusters. The
sample-to-sample deviations of these observables are thus predicted to present
power-law dependences with time, with exponents values that can be
determined from the Family-Vicsek exponents $\beta$ (growth exponent)
and $z$ (dynamical exponent).

We have tested our predictions against several different growth systems: random
deposition in two different versions (see Appendix \ref{sec:rd}), first-passage
percolation clusters and random metrics isochrones (both belonging to
the KPZ universality class), and shadowing dynamics (which does not).
In all the considered cases, the predictions met the actual scaling found in the 
simulations.

We have also addressed the full histogram of the sample-to-sample fluctuations of these global
variables. Some of them, such as the radius, area and length, are seen to be Gaussian. 
However, and in analogy to the case of KPZ growth in a band geometry \cite{Foltin.94,Antal.02,Halpin.15}, 
the histogram of the width or roughness is not Gaussian, and this remains beyond our present scaling arguments. 
The center-of-mass displacement follows a $\chi^2$-distribution with $k=2$ degrees of freedom, as predicted.
Also, the sample-to-sample correlation coefficients between these magnitudes approach time-independent, 
limiting values, also as predicted.

In principle, our work enables alternative characterizations of the universality class in terms of exponent values,
for rough interfaces with an overall, circular symmetry, by employing sample-to-sample fluctuations of global 
magnitudes associated to the clusters. Indeed, it is possible to obtain both the growth and the dynamical exponents using the
fluctuations in two complementary global magnitudes, such as the average radius and the total length. Methodologically, 
this may turn out to be advantageous in the analysis of e.g.\ experimental and/or simulation data.


\begin{acknowledgments}
We would like to thank Pedro Córdoba-Torres, Silvio C.\ Ferreira,
Olivier Pierre-Louis, and Kazumasa A.\ Takeuchi for very useful
discussions. This work was partially supported by Ministerio de
Ciencia e Innovaci\'on (Spain), Agencia Estatal de Investigaci\'on
(AEI, Spain, 10.13039/501100011033), and European Regional Development
Fund (ERDF, A way of making Europe) through Grants
Nos.\ PID2019-105182GB-I00 and PID2021-123969NB-I00, and by 
Comunidad de Madrid (Spain) under the Multiannual Agreement with UC3M 
in the line of Excellence of University Professors (EPUC3M14 and 
EPUC3M23), in the context of the V Plan Regional de Investigaci\'on 
Cient\'{\i}fica e Innovaci\'on Tecnol\'ogica (PRICIT). I.\ A.\ D.\ acknowledges funding from UNED through an FPI
scholarship.. D.\ V.\ 
acknowledges funding from Ministerio de Ciencia e Innovaci\'on trough 
FPI scolarship No.\ PRE2019-088226.
\end{acknowledgments}

\FloatBarrier


\appendix
\section{Random deposition}
\label{sec:rd}

The simplest growth model is, indeed, random deposition (RD), which
always yields $\beta=1/2$ \cite{Barabasi}. In a circular framework, we
may consider two flavors of the RD class (additional formulations are
possible, see e.g.\ \cite{Escudero.11} and related references), depending 
on the discretization scheme and the treatment of the UV-cutoff. In what we
will call model RD-1, we set up a fixed angular discretization with an
UV-cutoff $\Delta\theta=2\pi/N$, where $N$ is the number of points.
Now, the interface is described by a set of $N$ radial values,
$\{r_i\}_{i=1}^N$, with $r_i=r(\theta_i)$. At each time-step, we allow
each $r_i$ to grow independently of the others. In practice, we are
imposing that each wedge $\Delta\theta$ remains completely correlated.
Therefore, $\xi \sim t$, i.e. the correlated patches grow as fast as
the interface itself and $\zeta=1$.

\begin{figure}[h]
  \includegraphics[width=4cm]{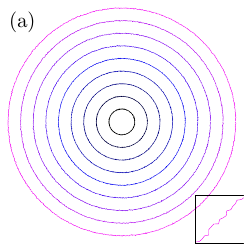}
  \hskip 0.5cm
  \includegraphics[width=4cm]{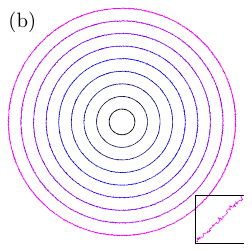}
  \caption{Profiles for the random deposition models discussed in
    Appendix \ref{sec:rd}: (a) Model RD-1 has a fixed angular cutoff;
    (b) Model RD-2 has an adaptive cutoff, thus the number of points
    along the interface grows witht time.}
  \label{fig:prof_rd}
\end{figure}

Model RD-2, on the other hand, includes a UV cutoff for length instead of
an angular one \cite{Rodriguez.11, Santalla.14, Santalla.15}. Therefore, the length of the correlated patches remains 
time independent, and $\zeta=0$. The differences between both RD models can be
seen in the profiles shown in Fig.\ \ref{fig:prof_rd}.

\begin{figure*}[!t]
  \includegraphics[width=8cm]{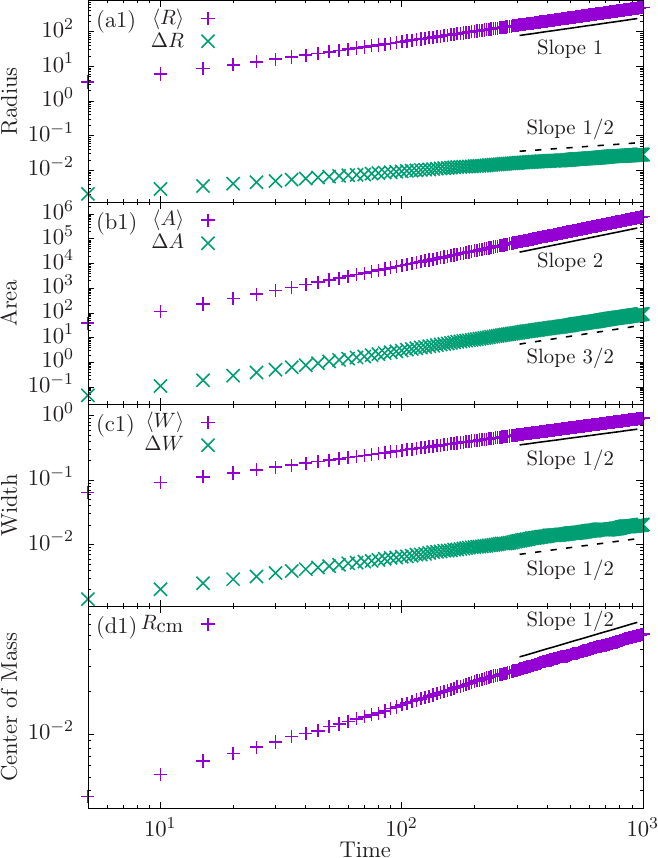}
  \hskip 1.5cm
  \includegraphics[width=8cm]{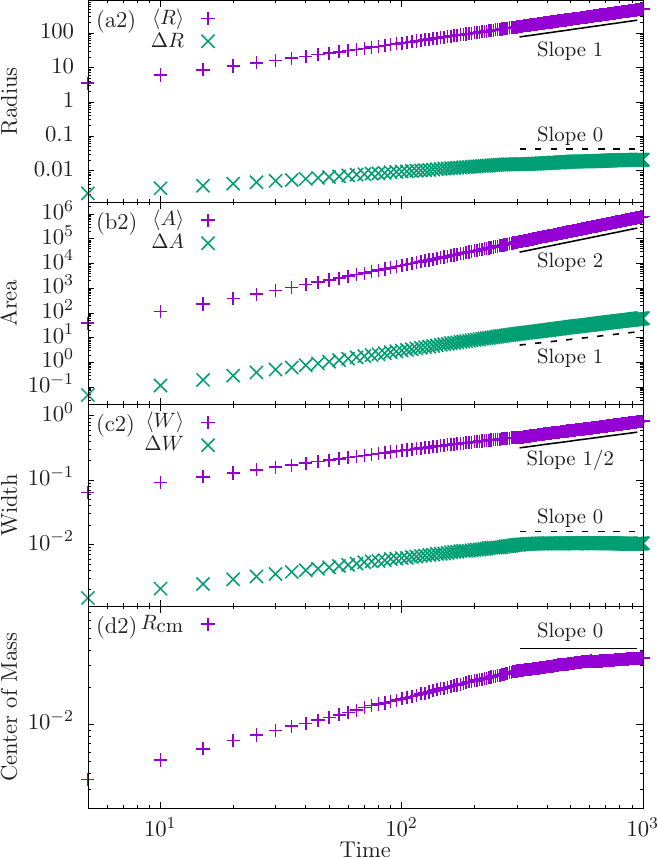}
  \caption{Time evolution of the average and sample-to-sample deviation 
    of global magnitudes from numerical simulations of model RD-1 in panels 
    (a1) to (d1), and of model RD-2 in panels (a2) to (d2). The straight lines in each panel represent the corresponding theoretical 
    expectation, see 
    Table \ref{table:rd}. The largest deviations between the numerical data
    and our scaling predictions are for the radial deviations [panel (a2)] and the CM 
    displacement [panel (d2)] of the RD-2 model. In both cases the expected exponent 
    is zero, while the measured value is approximately $0.13$.}
  \label{fig:rd}
\end{figure*}

The predictions for the sample-to-sample fluctuations of global
magnitudes vary for the two models. We discard the cluster length,
because our calculation assumed that the interface was smooth enough
at the cutoff scale, which is not the case here. The remaining scaling
exponents are shown in Table \ref{table:rd}, and have been measured in 
the numerical simulations shown in Fig.\
\ref{fig:rd}. In our simulations we have run $1000$ samples with a
growth velocity $v=1$, $\Delta t=0.01$, and unit adaptive UV-cutoff 
for the RD-2 model. The largest discrepancy between the theoretically
expected exponents and those measured in the simulations are found in model RD-2
for the deviations of the average radius and for the CM deviation; in both cases
we expect a zero exponent value but we measure $0.13$ approximately, 
possibly due to limitations in our longest simulation times. 
Other than this, the predictions seem accurate.

\begin{table}[!t]
 \begin{tabular}{@{}c|c|c|c|c|c}
      Observable & Scaling exponent & RD-1 & RD-2 \\
      \hline
      $\Delta R$ & $\beta+(\zeta-1)/2$ & 1/2 & 0\\
      $\Delta A$ & $\beta+(\zeta+1)/2$ & 3/2 & 1 \\
      $\Delta W$ & $\beta+(\zeta-1)/2$ & 1/2 & 0 \\
      $R_\CM$    & $\beta+(\zeta-1)/2$ & 1/2 & 0\\
      \hline
    \end{tabular}
    \caption{Scaling behavior of different geometric global
      observables for the RD-1 and RD-2 models. The growth exponent
      $\beta=1/2$ for both models, but $\zeta=1$ for RD-1 and
      $\zeta=0$ for RD-2.}
\label{table:rd}

\end{table}


\end{document}